\documentclass[12pt,superscriptaddress,aps,prb]{revtex4-1}

\usepackage[hmargin=2.0cm,vmargin=2.3cm]{geometry}
\usepackage{textcomp}

\usepackage{graphicx}

\usepackage{textpos}

\usepackage{amssymb}

\usepackage{amsmath}

\usepackage{txfonts}

\newcommand{\ket}[1]{\left|#1\right\rangle}

\newcommand{\Ca}{$^{40}$Ca$^{+}$}
\newcommand{\Ba}{Ba$^{+}$}

\begin{document}

\title{A Michelson-Morley Test of Lorentz Symmetry for Electrons}

\author{T. Pruttivarasin}
\affiliation{Department of Physics, University of California, Berkeley, California 94720, USA}
\affiliation{Quantum Metrology Laboratory, RIKEN, Wako, Saitama 351-0198, Japan}
\author{M. Ramm}
\affiliation{Department of Physics, University of California, Berkeley, California 94720, USA}
\author{S. G. Porsev}
\affiliation{Department of Physics and Astronomy, University of Delaware, Newark, Delaware 19716, USA}
\affiliation{Petersburg Nuclear Physics Institute, Gatchina, Leningrad District 188300, Russia}
\author{I. I. Tupitsyn}
\affiliation{Department of Physics, St. Petersburg State University, Ulianovskaya 1,
Petrodvorets, St.Petersburg 198504, Russia}
\author{M. Safronova}
\affiliation{Department of Physics and Astronomy, University of Delaware, Newark, Delaware 19716, USA}
\affiliation{Joint Quantum Institute, National Institute of Standards and Technology and the University of Maryland, College Park,
Maryland, 20742, USA}
\author{M. A. Hohensee}
\affiliation{Department of Physics, University of California, Berkeley, California 94720, USA}
\affiliation{Lawrence Livermore National Laboratory, Livermore, California 94550, USA}
\author{H. H\"affner}
\affiliation{Department of Physics, University of California, Berkeley, California 94720, USA}

\date{\today}

\maketitle


\textbf{All evidence so far suggests that the absolute spatial orientation of an
experiment never affects its outcome. This is reflected in the Standard Model of physics by requiring all particles and fields to be invariant under Lorentz transformations. The most well-known test of this important cornerstone of physics are Michelson-Morley-type experiments\cite{MM, Herrmann2009,Eisele2009} verifying the isotropy of the speed of light. Lorentz symmetry also implies that the kinetic energy of an electron should be independent of the direction of its velocity, \textit{i.e.,} its dispersion relation should be isotropic in space. In this work, we search for violation of Lorentz symmetry for electrons by performing an electronic analogue of a Michelson-Morley experiment. We split an electron-wavepacket bound inside a calcium ion into two parts with different orientations and recombine them after a time evolution of 95~ms. As the Earth rotates, the absolute spatial orientation of the wavepackets changes and anisotropies in the electron dispersion would modify the phase of the interference signal. To remove noise, we prepare a pair of ions in a decoherence-free subspace, thereby rejecting magnetic field fluctuations common to both ions\cite{Roos2006}. After a 23 hour measurement, we limit the energy variations to $h\times 11$ mHz ($h$ is Planck's constant), verifying that Lorentz symmetry is preserved at the level of $1\times10^{-18}$. We improve on the Lorentz-violation limits for the electron by two orders of magnitude\cite{Hohensee2013c}. We can also interpret our result as testing the rotational invariance of the Coloumb potential, improving limits on rotational anisotropies in the speed of light by a factor of five~\cite{Herrmann2009,Eisele2009}. Our experiment demonstrates the potential of quantum information techniques in the search for physics beyond the Standard Model.}


Invariance under Lorentz transformations is a key feature of the Standard Model (SM), and as such is fundamental to nearly every aspect of modern physics. Nevertheless, this symmetry may be measurably violated, \emph{e.g.,} due to spontaneous symmetry breaking in fields with dynamics at experimentally inaccessible energy scales not explicitly treated by the SM~\cite{KosteleckySamuel1989}. Some theories that unify gravitation and the SM assert that Lorentz symmetry is only valid at large length scales~\cite{Horava,Pospelov}. Other models suggest that strong Lorentz-violation at the Planck scale might be custodially suppressed by supersymmetry. In such scenarios, improved constraints on Lorentz-violation at low energy can be used to set an upper bound on the supersymmetric energy scale of on the order of 100~TeV (ref.~\citenum{LiberatiMattingly2012}). Therefore, precision tests of Lorentz symmetry complement direct probes of high energy physics being carried out at the Large Hadron Collider.

We analyse Lorentz-violation in the context of a phenomenological framework known as the Standard Model Extension (SME)~\cite{Colladay1997,Colladay1998}. The SME is an effective field theory that augments the SM Lagrangian with every possible combination of the SM fields that is not term-by-term Lorentz invariant, while maintaining gauge invariance, energy-momentum conservation, and Lorentz invariance of the total action~\cite{Colladay1997,Colladay1998}.  The SME can be used to describe the low-energy limit of many different theories which predict Lorentz-violation, and includes the SM as a limiting case. The SME thus provides a comprehensive framework for quantifying a wide range of Lorentz-violating effects, and is a flexible tool for consistently evaluating a wide variety of experiments~\cite{datatables}.

The SME allows for Lorentz-violation for all particles separately. However, to verify a particle's Lorentz symmetry, one must compare it to a reference system as only differences in their behaviors under Lorentz transformation are observable\cite{Colladay1998}. For instance, typical interpretations of Michelson-Morley-experiments testing Lorentz-violation of photons assume that the length of the interferometer arms are invariant under rotations. As the length of interatomic bonds depends on the electron's dispersion relation\cite{Mueller2003,Mueller2005}, those interpretations can be said to assume that Lorentz symmetry for electrons (and nuclei making up the interferometer arms) holds unless a second distinct reference system is used\cite{Mueller2005}. For our experiment, it is more natural to use light as a reference and assume that photons obey Lorentz symmetry. However, it is important to keep in mind that an experimental signature of the Lorentz-violation considered here can equally be attributed to Lorentz-violation of electrons as well as to that of photons (see Methods). 

The electronic Lorentz-violation of interest manifests via a modified quantum-electrodynamics Lagrangian:
\begin{align}
\mathcal{L} = \frac{1}{2}i\bar{\psi}(\gamma_\nu+c_{\mu\nu}\gamma^\mu)\stackrel{\leftrightarrow}{D^\nu}\psi-\bar{\psi}m_\text{e}\psi, \label{eq:SME_lagrangian}
\end{align}
where $m_\text{e}$ is the electron mass, $\psi$ is a Dirac spinor, $\gamma^\mu$ are the Dirac matrices, $\bar{\psi}\stackrel{\leftrightarrow}{D^\nu}\psi \equiv \bar{\psi} D^\nu \psi-\psi D^\nu \bar{\psi}$ with $D^\nu$ being the covariant derivative, and finally $c_{\mu\nu}$ is a symmetric tensor describing Lorentz-violation\cite{Colladay1997,Colladay1998}.  Since $c_{\mu\nu}$ is frame dependent, we uniquely specify its value in the Sun-centred, celestial-equatorial frame (SCCEF), \textit{i.e.} the Sun's rest frame. Time-dependent Lorentz transformations due to the Earth's motion transform $c_{\mu\nu}$ in the SCCEF to the time-dependent values in the local laboratory frame on the Earth. Hence, the contribution of $c_{\mu\nu}$ to any laboratory-frame observable will vary in time. 

For us, the important consequence of electronic Lorentz-violation is the dependence of an electron's energy on the direction of its momentum. For an atomically bound electron with momentum $\mathbf{p}$, the Lagrangian in Eq.\ (\ref{eq:SME_lagrangian}) results in a small energy shift that depends on the direction of the electron's momentum described by the effective Hamiltonian~\cite{Kostelecky1999}
\begin{align}
\delta\mathcal{H} = -C^{(2)}_0\frac{(\mathbf{p}^2-3p^2_z)}{6m_\text{e}},\label{eq:SME_hamiltonian}
\end{align}
where $C_0^{(2)}$ contains elements in $c_{\mu\nu}$ in the laboratory frame and $p_z$ is the component of electron momentum along the quantisation axis which is fixed in the laboratory. The energy shift depends on how the total momentum $\mathbf{p}$ is distributed among the three spatial components. As the Earth rotates, $C_0^{(2)}$ varies in time, resulting in a time variation of the electron's energy correlated with the Earth's motion.

\begin{figure}
\includegraphics[width=0.5\textwidth]{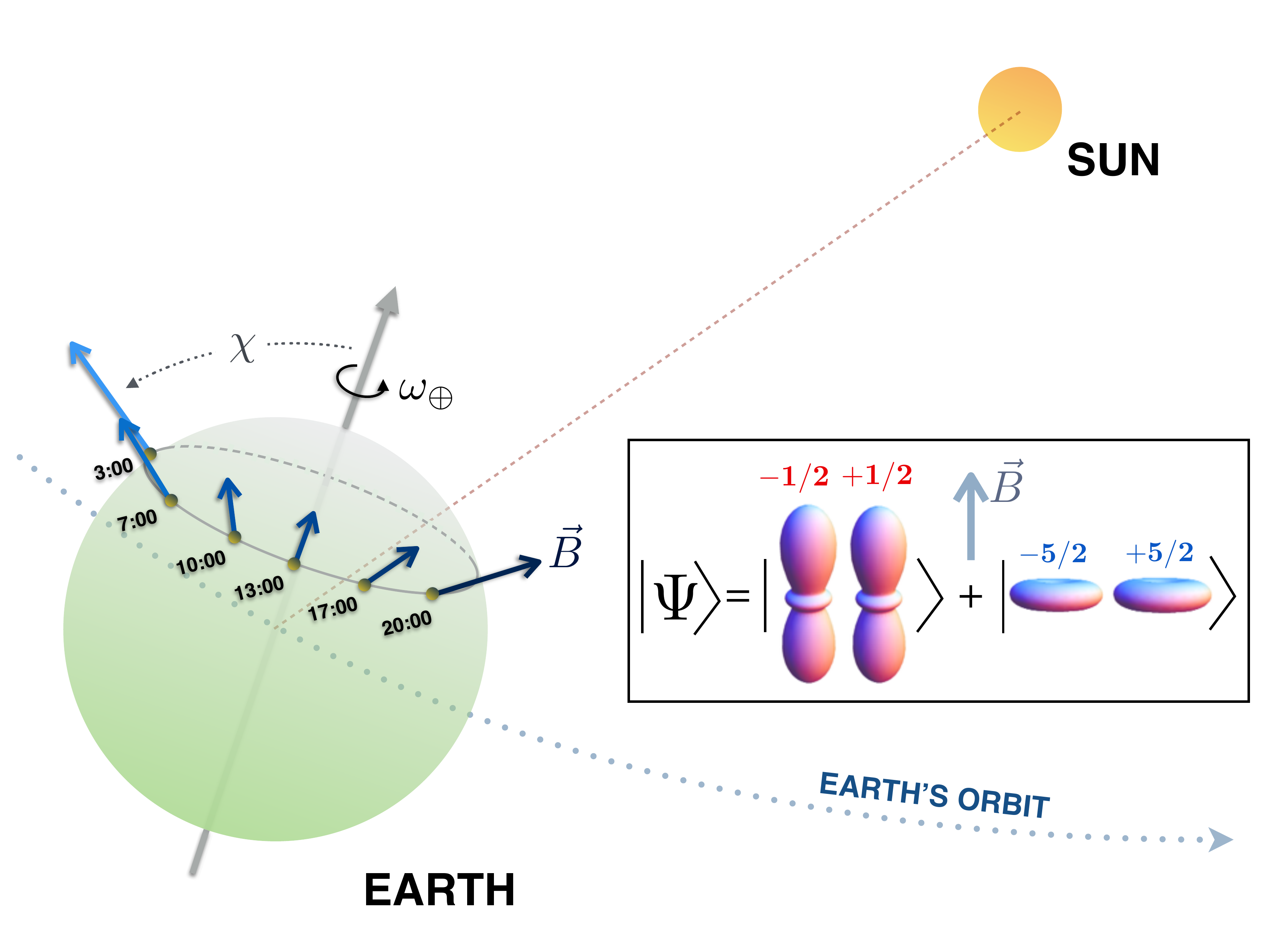}
\caption{\textbf{Rotation of the quantisation axis of the experiment with respect to the Sun as the Earth rotates.} We apply a magnetic field ($\vec{B}$) of 3.930~G vertically in the laboratory frame to define the quantisation axis of the experiment. As the Earth rotates with an angular frequency given by $\omega_\oplus=2\pi/23.93 \text{ h}$, the orientation of the quantisation axis and consequently that of the the electronic wavepacket (as shown in the inset) changes with respect to the Sun's rest frame. The angle $\chi\sim52.1$\textdegree~is the colatitude of the experiment.}
\label{fig:globe}
\end{figure}

To probe Lorentz-violation, we perform the electronic analogue of a Michelson-Morley experiment by interfering atomic states with anisotropic electron momentum distributions aligned along different directions, such as available in the $^2$D$_{5/2}$ manifold of~\Ca. We trap a pair of~\Ca~with a separation of $\sim$ 16 $\mu$m in a linear Paul trap, and define the quantisation axis by applying a static magnetic field of 3.930~G vertically. The direction of this magnetic field changes with respect to the Sun as the Earth rotates, resulting in a rotation of our interferometer (see Figure\ \ref{fig:globe}).

We calculate the hypothetical energy shift of~\Ca~in the $^2$D$_{5/2}$ manifold according to Eq.\ (\ref{eq:SME_hamiltonian}):
\begin{align}
\frac{\Delta E_\text{LLI}}{h} = [(2.16 \times 10^{15})-(7.42 \times 10^{14})\cdot m_J^2]\cdot C_0^{(2)} \text{ ~~(Hz)},\label{eq:energy_shift}
\end{align}
where $m_J$ is the magnetic quantum number (see Supplementary Information). To obtain maximum sensitivity to Lorentz-violation, we monitor the energy difference between the state $\ket{\pm5/2}\equiv\ket{^2\text{D}_{5/2};m_J=\pm5/2}$ and $\ket{\pm1/2}\equiv\ket{^2\text{D}_{5/2};m_J=\pm1/2}$ using a Ramsey-type interferometric scheme. To reject magnetic field noise which is the main source of decoherence, we create a product state $\ket{\Psi^\text{P}} = \frac{1}{2}\left(\ket{-1/2}+\ket{-5/2}\right)\otimes(\ket{+1/2}+\ket{+5/2})$ by applying a series of $\pi/2$ and $\pi$ pulses on the S-D transition to both ions. Under common noise induced by a fluctuating magnetic field, the product state rapidly dephases to a mixed state that contains a decoherence-free entangled state $\ket{\Psi^{R}}\equiv\frac{1}{\sqrt{2}}(\ket{-5/2,+5/2}+\ket{-1/2,+1/2})$ with 50\% probability\cite{Chwalla2007}. This entangled state time-evolves freely according to 
\begin{align}
\ket{\Psi^{R}(t)} = \frac{1}{\sqrt{2}}\left(\ket{-5/2,+5/2}+e^{i(\Delta E_{R} t/\hbar+\phi_{R})}\ket{-1/2,+1/2}\right)
\end{align}
where $\Delta E_{R}$ is the energy difference between the state $\ket{-5/2,+5/2}$ and $\ket{-1/2,+1/2}$, and $\phi_{R}$ is a phase offset. The remaining components of the mixed state, which are the state $\ket{+1/2,-5/2}$ and $\ket{-1/2,+5/2}$, each with 25\% probability, are time-independent.

\begin{figure}
\includegraphics[width=0.5\textwidth]{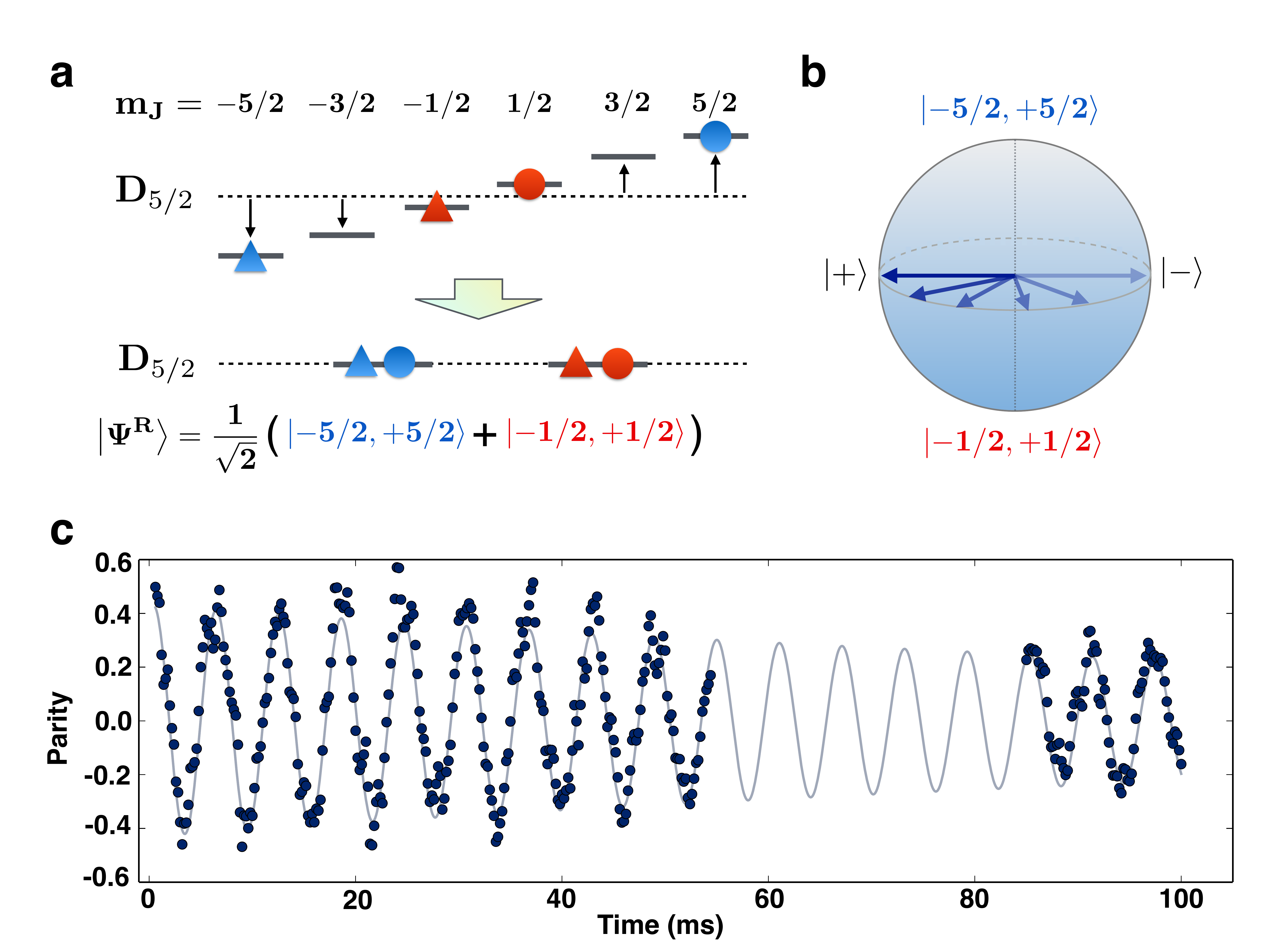}
\caption{\textbf{Oscillation of the decoherence-free state.} \textbf{a,} A combination of different magnetic sub-levels of the first (denoted by $\bullet$) and second (denoted by $\blacktriangle$) \Ca~ions in the $^2$D$_{5/2}$ manifold forms a decoherence-free state $\ket{\Psi^{R}}\equiv\frac{1}{\sqrt{2}}\left(\ket{-5/2,+5/2}+\ket{-1/2,+1/2}\right)$. Blue and red colours indicate pairing of the single ion states in each component of $\ket{\Psi^{R}}$. \textbf{b,} Time evolution of the state $\ket{\Psi^{R}(t)}$ represented by a trajectory on a Bloch sphere where the poles are the $\ket{-5/2,+5/2}$ and $\ket{-1/2,+1/2}$ states. The state $\ket{\Psi^{R}(t)}$ oscillates back and forth between the even-odd parity basis states, $\ket{\pm}$, as given in Eq.\ (\ref{eq:parity_oscillation}). \textbf{c,} Oscillation of a product state that dephases into a mixed state that contains an entangled state $\ket{\Psi^{R}}$ with 50\% probability. Each data point is taken with 200 repetitions of the Ramsey-type experimental cycle shown in Fig.\ \ref{fig:measurement}a. The grey solid line is a fit to the Ramsey fringe function with an oscillation frequency of $164.9\pm0.1$ Hz. The fit yields a decay constant of $155\pm17$ ms, which is substantially shorter than the value expected from the lifetime of the $^2$D$_{5/2}$ state of \Ca. We attribute the loss of coherence to the heating rate of the ion trap of $\sim0.2$ quanta/ms which degrades the quality of the analysis pulses for long Ramsey interrogation times.}
\label{fig:bloch}
\end{figure}

In Figure\ \ref{fig:bloch}, we illustrate the dynamics of the state $\ket{\Psi^R}$. By expressing the state in the even-odd parity basis, $\ket{\pm} = \frac{1}{\sqrt{2}}(\ket{-5/2,+5/2}\pm\ket{-1/2,+1/2})$, the time evolution $\ket{\Psi^R(t)}$ can be written as
\begin{align}
\ket{\Psi^R(t)} = \frac{1}{\sqrt{2}}\left((1+e^{i(\Delta E_{R} t/\hbar+\phi_{R})})\ket{+}+(1-e^{i(\Delta E_{R} t/\hbar+\phi_{R})})\ket{-}\right).\label{eq:parity_oscillation}
\end{align}
We interpret the trajectory of $\ket{\Psi^R(t)}$ to be along the equator of the Bloch sphere as shown in Figure\ \ref{fig:bloch}b. The state $\ket{\Psi^R(t)}$ oscillates back and forth between the state $\ket{+}$ and $\ket{-}$ with frequency $f_R = \Delta E_R/h$. To readout the ion state in the $\ket{\pm}$ basis, we apply a series of $\pi$ and $\pi/2$ pulses on the S-D transition to both ions followed by an electron shelving readout scheme\cite{Roos2006}. The difference between the probability $P_+$ and $P_-$ for the ions to be in the state $\ket{+}$ and $\ket{-}$, respectively, yields an oscillating signal given by $P = P_+ - P_- = \cos{(\Delta E_{R} t/\hbar + \phi_R)}$, as shown in Figure\ \ref{fig:bloch}c. 

We are interested in the variations of the energy difference between the $\ket{\pm5/2,\mp5/2}$ and $\ket{\pm1/2,\mp1/2}$ states due to Lorentz-violation. However, linear Zeeman shifts from a residual magnetic field gradient, quadratic Zeeman shifts, electric quadrupole shifts from an electric field gradient, and ac Stark shifts from oscillating trapping fields also affect the energy difference\cite{Chou2010,Madej2012}. The contributions from the magnetic field gradient on the order of 100~Hz have opposite signs for the state $\ket{\Psi^R}$ and its mirrored counterpart, $\ket{\Psi^{L}}\equiv\frac{1}{\sqrt{2}}\left(\ket{+5/2,-5/2}+\ket{+1/2,-1/2}\right)$. We can correct for this contribution to the oscillation signal by taking the average frequency $\bar{f} = (f_R+f_L)/2$. The remaining effects (except for Lorentz-violation), are energy shifts on the order of only a few Hertz and are also directly related to external electromagnetic fields in the proximity of the ions. We expect these fields to be stable to the $10^{-3}$ level in a day and the associated variations are on the few mHz level and below. Moreover, we independently measure these fields using the ions themselves as a probe (see Methods).

\begin{figure}
\includegraphics[width=0.5\textwidth]{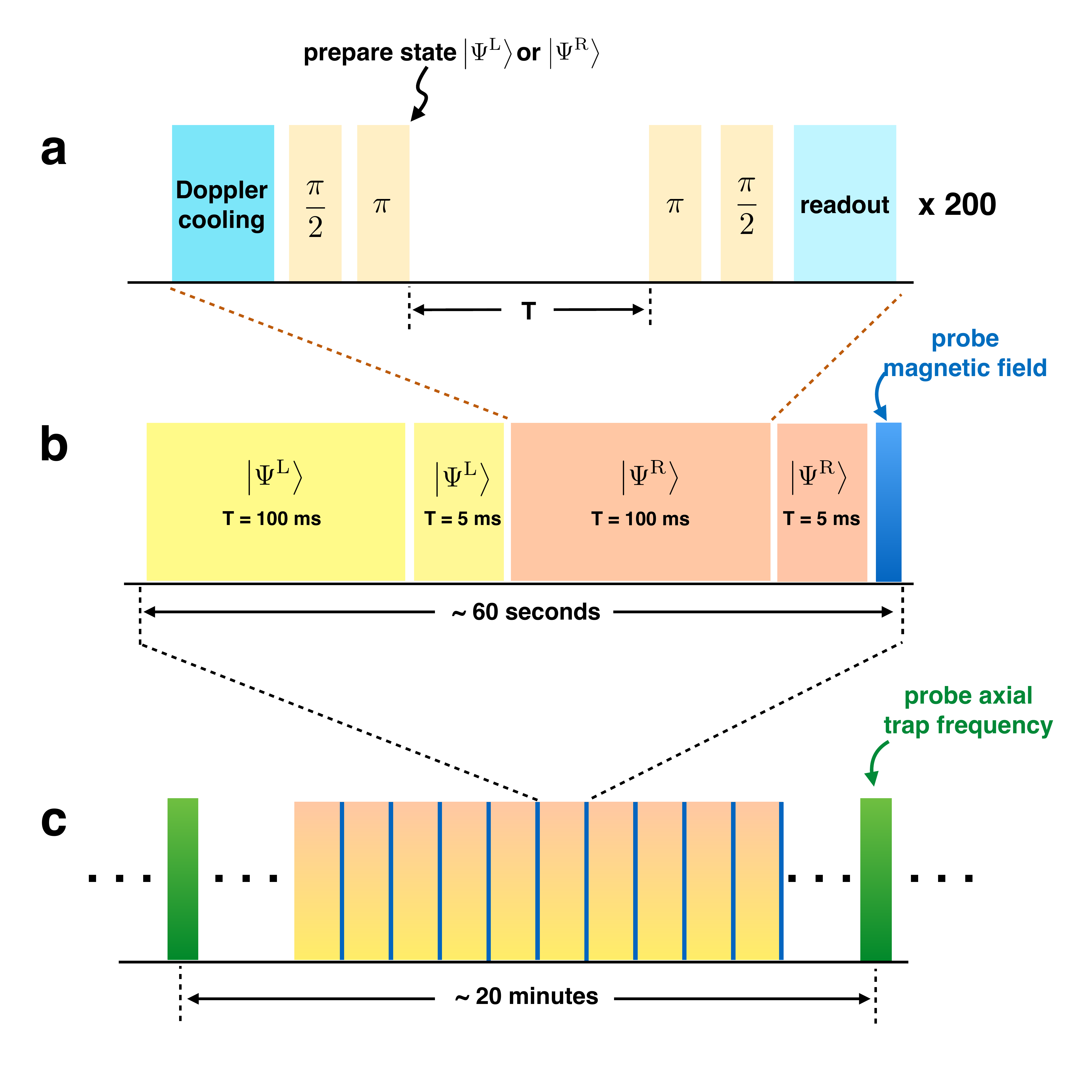}
\caption{\textbf{Outline of the experimental scheme.} \textbf{a,} The building block of our experiment is a Ramsey-type interferometric sequence. In each measurement cycle, we first perform Doppler cooling and optical pumping of the ions. Then, a series of $\pi/2$ and $\pi$ pulses on the S-D transition prepare the ions in a product state that dephases into a mixed state within 1 ms. This state contains an entangled state $\ket{\Psi^{L,R}}\equiv\frac{1}{\sqrt{2}}\left(\ket{\pm5/2,\mp5/2}+\ket{\pm1/2,\mp1/2}\right)$ with 50\% probability. Afterwards, the mixed state evolves freely for Ramsey duration $T$, before another series of $\pi$ and $\pi/2$ pulses, together with an electron shelving readout sequence, allows us to readout the state of the ions in the even-odd parity basis. This measurement cycle is repeated for 200 times for $\ket{\Psi^{L}}$ and $\ket{\Psi^{R}}$.  \textbf{b,} To correct for phase drifts in the preparation of $\ket{\Psi^{L,R}}$, we measure the difference in the oscillation signal between Ramsey durations of $100$ ms and $5$ ms. We then correct for the contribution of the magnetic field gradient by taking the average of the oscillation signals measured with state $\ket{\Psi^L}$ and $\ket{\Psi^R}$. At the end of this measurement block, we measure the magnetic field by performing spectroscopy on the S-D transition to correct for the quadratic Zeeman effect. Each grey data point in Fig.\ \ref{fig:allan}a is a result from one of these measurement blocks. \textbf{c,} We continuously repeat the measurement block during the course of the 23-hour long measurement. To correct for the electric quadrupole shift caused by the electric field gradient, we measure the axial trap frequency by performing spectroscopy on the S-D transition.
}
\label{fig:measurement}
\end{figure}

\begin{figure}
\includegraphics[width=0.5\textwidth]{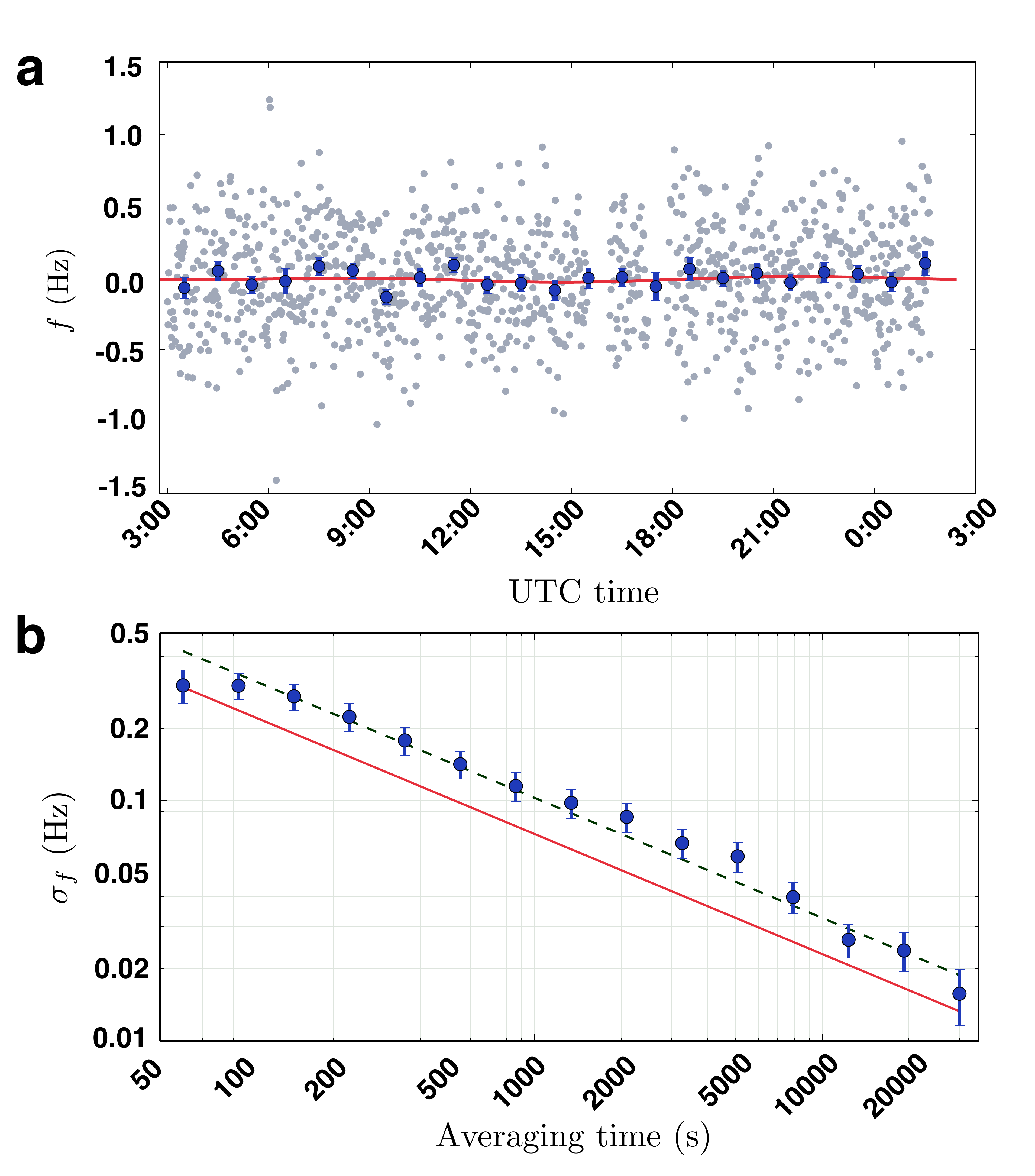}
\caption{\textbf{Frequency measurements for~\Ca.} \textbf{a,} The grey coloured data points represent frequency measurements of~\Ca~taken after each measurement block as shown in Fig.\ \ref{fig:measurement}b with contributions from the quadratic Zeeman shifts and electric quadrupole shifts subtracted out. (Gaps in the data points are due to a failure of the laser frequency stabilisation). We started the measurement at 3:00 UTC of April 19th, 2014, and continued for 23 hours. Dark blue points are obtained by binning of data from 60 minute time intervals. The errorbars represent the 1$\sigma$ standard error of the data points within the bin, where we scale the error by $\sqrt{\chi^2_\text{reduced}} = 1.3$ (obtained from the fit of the binned data to the model in Eq.\ (\ref{eq:model})). \textbf{b,} Allan deviation of the frequency measurement, $\sigma_f$, calculated from the unbinned data. The red solid line is the estimated quantum projection noise. The green dashed line is a fit to the data, showing a sensitivity to the ions' energy variation of $\sigma_f = 3.3\text{ Hz} /\sqrt{\tau}$, where $\tau$ is the averaging time. The steady downward trend indicates that we are still limited by statistical fluctuations rather than by correlated noise or systematics over the course of the measurement.}
\label{fig:allan}
\end{figure}

We measured the energy difference between the state $\ket{\pm5/2,\mp5/2}$ and $\ket{\pm1/2,\mp1/2}$ of \Ca~for 23 hours starting from 3:00 AM Coordinated Universal Time (UTC) on 19th April, 2014, by monitoring the oscillation signal of the ions with an effective Ramsey duration of 95 ms (see Methods). At the same time, we monitored the magnetic field and the electric field gradient using the ions themselves as a probe (see Figure \ref{fig:measurement}). 
We then used the measured values of the magnetic field and electric field gradient to correct for the quadratic Zeeman and electric quadrupole shifts. The resulting 23-hour frequency measurement is shown in Figure~\ref{fig:allan}. With 23 hours of averaging, we reach a sensitivity of the oscillation frequency of 11 mHz, limited by statistical uncertainties due to short term fluctuations. We then attribute any residual variation of the energy correlated with the Earth's rotation to Lorentz-violation. 

Lorentz transformations of $c_{\mu\nu}$ from the SCCEF to the laboratory frame results in the time-dependent energy shift due to Lorentz-violation given by
\begin{align}
\frac{\Delta E_\text{LLI}}{h} = A\cos(\omega_\oplus T) + B\sin(\omega_\oplus T) + C\cos(2\omega_\oplus T) + D\sin(2\omega_\oplus T), \label{eq:model}
\end{align}
where $\omega_\oplus =2\pi/23.93\text{ h}$ is the sidereal angular frequency of the Earth's rotation, $T$ is time since vernal equinox of 2014 and $(A,B,C,D)$ are parameters related to $c_{\mu\nu}$ in the SCCEF (see Supplementary Information). Fitting our data (Figure~/ref{fig:allan}) to Eq.\ (\ref{eq:model}) yields the limits of the $c_{\mu\nu}$ parameters, where we report in Table \ref{table:cparam} our results compared to existing limits. We improve the best direct measurements of the electron dispersion carried out by precision spectroscopy of dysprosium\cite{Hohensee2013c} by up to two orders of magnitude to a level of $1\times10^{-18}$. Recalling that our analysis assumed that the speed of light is constant, we can alternatively interpret our results as limits for Lorentz-violation for photons provided that Lorentz symmetry holds for electrons (see Methods). Doing so, we improve on the bounds for Lorentz symmetry set by photon-Michelson-Morley experiments\cite{Herrmann2009} by up to five times (see Table \ref{table:cparam}). 

Our experimental scheme is readily applicable to other trapped ion species. Further improvement can be achieved by increasing the Ramsey durations by utilising metastable states with significantly longer lifetime, such as 30~seconds for barium\cite{Iskrenova-Tchoukova2008}, or by using ions with higher sensitivity to Lorentz-violation, such as highly charged ions\cite{Safronova2014}. Additionally, by preparing a pure entangled state of the ions instead of a mixed state, one readily gains another factor of two in signal-to-noise ratio\cite{Roos2006}. Finally, we do not see any signature of limiting systematic effects and thus expect that future extensions of our experimental technique will yield much improved tests of Lorentz symmetry.

\clearpage

\clearpage

\clearpage

\clearpage

\clearpage

\begin{table}
\caption{\textbf{Limits on Lorentz-violation parameters $c_{\mu\nu}$.} Fitting our frequency measurements to the model in Eq.\ (\ref{eq:model}) yields the limits on Lorentz-violation parameters $c_{\mu\nu}$ in the SCCEF. All uncertainties for the uncorrelated combinations of $c_{\mu\nu}$ are 1$\sigma$ standard errors from the fit conservatively scaled with $\sqrt{\chi^2_\text{reduced}} = 1.3$. Assuming that Lorentz symmetry holds for electrons, our results improve on the existing limits set by a modern version of the classic Michelson-Morley experiment in ref.\ \citenum{Herrmann2009} by up to five times. Taking the alternative view where Lorentz symmetry of photons holds, we improve the bounds for the electron dispersion relation from ref. \citenum{Hohensee2013c} by up to two orders of magnitude. Note that we use the notation $c_{X-Y} = c_{XX}-c_{YY}$.}
\begin{tabular}{c c c c}
\\
\hline\hline
Parameters & New limits& \multicolumn{2}{c}{Existing limits} \\
 & & photon (ref.\ \citenum{Herrmann2009}) & electron (ref.\ \citenum{Hohensee2013c}) \\
\hline
-0.16c$_{X-Y}$+0.33c$_{XY}$-0.92c$_{XZ}$-0.16c$_{YZ}$& $0.1\pm1.0\times10^{-18}$&$-2.5\pm3.5\times10^{-18}$& $-0.9\pm1.0\times10^{-16}$\\
-0.04c$_{X-Y}$-0.32c$_{XY}$-0.35c$_{XZ}$+0.88c$_{YZ}$ & $2.4\pm7.4\times10^{-19}$&$-5.2\pm3.6\times10^{-18}$& $-0.9\pm6.5\times10^{-17}$\\
0.29c$_{X-Y}$-0.38c$_{XY}$-0.73c$_{XZ}$-0.48c$_{YZ}$ & $5.9\pm9.5\times10^{-19}$&$-0.6\pm3.8\times10^{-18}$& $-8.1\pm9.5\times10^{-17}$\\
-0.31c$_{X-Y}$-0.65c$_{XY}$+0.07c$_{XZ}$-0.69c$_{YZ}$ & $0.7\pm1.2\times10^{-18}$&$-2.6\pm3.8\times10^{-18}$& $-2.9\pm6.5\times10^{-17}$\\
\hline
\end{tabular}
\label{table:cparam}
\end{table}

\subsection*{Acknowledgements}
This work was supported by the NSF CAREER program grant \# PHY 0955650, NSF grant \# PHY 1212442 and \# PHY 1404156, and was performed under the auspices of the U.S. Department of Energy by Lawrence Livermore National Laboratory under Contract DE-AC52-07NA27344. We thank Holger M{\"u}ller for critical reading of the manuscript.

\subsection*{Author Contributions} 
H.H., M.H., and T.P. conceived the experiment. T.P. and M.R. carried out the measurements. S.G.P., I.I.T., and M.S. calculated the sensitivity of the energy to Lorentz-violation. T.P., M.H. and H.H. wrote the main part of the manuscript, S.G.P., I.I.T., and M.S. the Supplementary Information. All authors contributed to the discussions of the results and manuscript.

\subsection*{Author Information} The authors declare no competing financial interests.
Readers are welcome to comment on the online version of the paper. Correspondence
and requests for materials should be addressed to T.P. (thaned.pruttivarasin@riken.jp) and H.H. (hhaeffner@berkeley.edu).

\clearpage

\section*{Methods}

\subsection*{Mapping of Lorentz-violation between electrons and photons.}

While Lorentz symmetry, or local Lorentz invariance, requires that the laws of physics be the same in all coordinate systems in the group formed by Lorentz transformations, it does not restrict our initial choice of coordinates. As a result, some forms of Lorentz-violation cannot be unambiguously attributed to a single species of elementary particle without first specifying this coordinate choice. In particular, we can select our initial coordinates such that $c_{\mu\nu}$ (or its gauge field analog $k_{\mu\nu}$) vanishes at leading order for any single species of particle (or gauge field). This particle then becomes a Lorentz-covariant `yardstick' which other species can be compared against. In Eq.\ (\ref{eq:SME_lagrangian}), we use light as our yardstick, \textit{i.e.,} we measure space such that $x_i = c_i t$ with the speed of light $c_i$ constant in all three spatial directions $i$. Alternatively, we might use a coordinate system for which Lorentz symmetry is preserved for electrons. In this case, Lorentz-violation would manifest itself by breaking the rotational symmetry of the Coulomb force, yielding the same measurable energy shift as in the previous case. 

To transform between both views, we neglect contributions of the nucleus to the Lorentz-violation signal for two reasons. First, the quadrupole moment of the doubly-magic \Ca-nucleus is expected to vanish.  Secondly, the violations Lorentz symmetry for nucleon constituents have been constrained to $10^{-26}$ for protons\cite{Wolf2006} and $10^{-29}$ for neutrons\cite{Romalis2011}. The Lorentz-violation for electromagnetic fields in the SME is then given by the $\tilde{\kappa}$ parameters (which are functions of $k_{\mu\nu}$) in the following Lagrangian\cite{Mewes2002}:
\begin{align}
\mathcal{L} &= \frac{1}{2}\left[(1+\tilde{\kappa}_\text{tr})|\vec{E}|^2-(1-\tilde{\kappa}_\text{tr})|\vec{B}|^2\right] + \frac{1}{2}\left[\vec{E}\cdot\tilde{\kappa}_{e-}\cdot\vec{E}-\vec{B}\cdot\tilde{\kappa}_{e-}\cdot\vec{B}\right]+\vec{E}\cdot\tilde{\kappa}_{o+}\cdot\vec{B},
\end{align}
where $\tilde{\kappa}_\text{tr}$ is a scalar and $\tilde{\kappa}_{e-}$ is a $3\times3$ traceless symmetric matrix and $\tilde{\kappa}_{o+}$ is an antisymmetric matrix. By means of a coordinate transformation, these $\tilde{\kappa}$ parameters can be mapped to elements in the $c_{\mu\nu}$ matrix in Eq.~(\ref{eq:SME_lagrangian}). The parameter relevant to our experiment is $\tilde{\kappa}_{e-}$ which characterises anisotropy of the speed of light. The mapping between $\tilde{\kappa}_{e-}$ and $c_{\mu\nu}$ is given explicitly by\cite{Bailey2004}
\begin{align}
c_{X-Y}\equiv  c_{XX}-c_{YY}  &= \frac{1}{2} (\tilde{\kappa}^{XX}_{e-} - \tilde{\kappa}^{YY}_{e-})\label{ckappa1}\\
c_{XY} &= \frac{1}{2} \tilde{\kappa}^{XY}_{e-}\\
c_{XZ} &= \frac{1}{2} \tilde{\kappa}^{XZ}_{e-}\\
c_{YZ} &= \frac{1}{2} \tilde{\kappa}^{YZ}_{e-}\label{ckappa4}.
\end{align}
The best existing limits on $\tilde{\kappa}^{XY}_{e-}, \tilde{\kappa}^{XZ}_{e-}$, $\tilde{\kappa}^{YZ}_{e-}$ and $(\tilde{\kappa}^{XX}_{e-}-\tilde{\kappa}^{YY}_{e-})$ are given in ref. \citenum{Herrmann2009}. Using the expression in Eq.\ (\ref{ckappa1}) to (\ref{ckappa4}), we can compare our result with the bounds for Lorentz-violation of photons, as shown in Table \ref{table:cparam}.

\subsection*{Experimental setup.} 

We trap a pair of \Ca~ions in a linear Paul trap with an interelectrode distance of 1.0 mm. We apply a radio frequency (rf) voltage of  $\sim500$ V$_\text{pp}$ to each pair of the rf electrodes. One pair of the electrodes is driven 180 degrees relative to the other pair. With $\sim$ 4 V dc applied to the endcaps, we obtain trap frequencies of 2.2 and 2.0 MHz in the radial directions and 210 kHz in the axial direction. The axial direction is aligned horizontally in the laboratory frame. To define a quantisation axis, we apply a static magnetic field of 3.930~G vertically (45 degrees with respect to both radial directions of the trap) using a coil. Additionally, we use another magnetic coil to compensate for residual magnetic field gradient along the axial direction. 

Two independent 729 nm laser light beams in the vertical direction drive $\pi$ and $\pi/2$ pulses on the S$_{1/2}$-D$_{5/2}$ transition on each ion separately. Both beams are derived from a laser stabilised to a high finesse optical cavity to better than 100~Hz.  Another beam path addressing both ions in the horizontal direction (45 degrees with respect to the axial direction) is used for Doppler cooling (397 nm and 866 nm) and repumping for the D$_{5/2}$ state (854 nm). We perform all laser light switching and frequency shifting using acousto-optical modulators (AOMs) in a double-pass configuration. We generate all rf voltages supplied to the AOMs using direct-digital-synthesiser (DDS) chips from Analog Devices$^\text{\textregistered}$ (AD9910). The timing in the experimental sequence is controlled by a field-programmable-gate-array (FPGA) module XEM6010 from Opal Kelly$^\text{\textregistered}$. We characterise the stability of the on-board crystal oscillator using a frequency counter (Agilent$^\text{\textregistered}$ 53210A). The clock stability is measured to be at the level of $4\times 10^{-7}$, which translates to better than $5~\mu\text{Hz}$ stability in the oscillation signal of the measurement of Lorentz-violation.

\subsection*{Measurement scheme.}

The experimental sequence is shown in Figure \ref{fig:measurement}. We measure four independent oscillation signals for the two states $\ket{\Psi^\text{L}}\equiv\frac{1}{\sqrt{2}}\left(\ket{+5/2,-5/2}+\ket{+1/2,-1/2}\right)$ and $\ket{\Psi^\text{R}}\equiv\frac{1}{\sqrt{2}}\left(\ket{-5/2,+5/2}+\ket{-1/2,+1/2}\right)$, each with both short ($T_\text{short}=5\text{ ms}$) and long ($T_\text{long}=100\text{ ms}$) Ramsey duration (see Figure \ref{fig:measurement}b). Within each measurement block in Figure \ref{fig:measurement}b, the order in which we perform Ramsey spectroscopy for each state and Ramsey duration is randomised to average out systematic noise that might coincide with the period ($\sim$ 60 seconds) of the measurement block.

In general, the oscillation signal has the form $S(t) = \mathcal{A}\cos(2\pi f t +\phi_\text{offset}+\phi_\text{laser})+\mathcal{B}$, where $\mathcal{A}$ is the amplitude, $\mathcal{B}$ is a possible offset to the overall level of the signal, $f$ is the oscillation frequency, $\phi_\text{offset}$ is the phase offset and $\phi_\text{laser}$ is an additional phase that we can control by changing the phase of the 729 nm laser light (through the rf signal supplied to the AOM for each beam path) that drives $\pi$ and $\pi/2$ pulses on the S$_{1/2}$-D$_{5/2}$ transition of the ions.

For a given state and Ramsey duration, the Ramsey interferometric cycle shown in Figure \ref{fig:measurement}a is repeated for 200 times. To cancel out drifts in the offset of the signal, $\mathcal{B}$, we perform the first 100 cycles of the Ramsey sequence with the phase of the laser light given by $\phi_\text{laser}$ and the next 100 cycles with the phase of the laser light given by $\phi_\text{laser}+\pi$. We then take the difference between these two signals, $(S(\phi_\text{laser}) - S(\phi_\text{laser}+\pi))/2 = \mathcal{A}\cos(2\pi f t +\phi_\text{offset}+\phi_\text{laser})$, which does not depend on $\mathcal{B}$.

For a fixed Ramsey duration $T$, the oscillation signal $S(T) = \mathcal{A}\cos(2\pi f T +\phi_\text{offset}+\phi_\text{laser})$ is most sensitive to variation in the oscillation frequency, $f$, when the signal crosses zero, \textit{i.e.} when $2\pi f T +\phi_\text{offset}+\phi_\text{laser} = \pi/2$. We make sure that the oscillation signal remains close to zero by adding the phase correction calculated from the oscillation signal: $\delta\phi = \cos^{-1}\left(\frac{S(T)}{\mathcal{A}}\right)-\frac{\pi}{2}$ to the phase of the laser light, $\phi_\text{laser}$. The long term measurement of the variation in the oscillation frequency, $\delta f$, is then derived from the phase correction data using $\delta \phi = 2\pi T \delta f$.

In addition to the change in the oscillation frequency, any change in $\phi_\text{offset}$ in the state preparation affects the phase correction: $\delta \phi = 2\pi T \delta f + \delta\phi_\text{offset}$. To correct for a contribution from this phase offset, we use signals from two Ramsey durations ($T_\text{short}=5\text{ ms}$ and $T_\text{long}=100\text{ ms}$) and calculate the difference between the phase corrections: $\delta \phi_\text{long} - \delta \phi_\text{short} = 2\pi (T_\text{long}-T_\text{short}) \delta f$. The oscillation frequency for the state $\ket{\Psi^\text{L,R}}$ is given by $\delta f_\text{L,R} = \left[(\delta \phi_\text{long} - \delta \phi_\text{short})/2\pi(T_\text{long}-T_\text{short})\right]_\text{L,R}$ where the effective Ramsey duration is $T_\text{long}-T_\text{short} = 95\text{ ms}$.

While the linear Zeeman effects from a magnetic field common to both ions drops out, the linear Zeeman effect due to a magnetic field gradient does not cancel. To remove these variations, we take the average frequency $\delta\bar{f} = (\delta f_\text{L}+\delta f_\text{R})/2$ of the states $\ket{\Psi^\text{L}}$ and $\ket{\Psi^\text{R}}$, which now contains only contributions from the electric quadrupole shift, quadratic Zeeman shift, ac Stark shifts from oscillating trapping fields and shifts from Lorentz-violation.

We characterise the effect of the electric quadrupole shift by measuring the oscillation frequency $\delta\bar{f}$ as a function of the electric field gradient by changing the axial trap frequency. For our experimental setup, we obtain $\delta\bar{f} = [4.0(8) \text{ (Hz mm$^2$/V)} \cdot E' +8.9(8) \text{ (Hz)}]$, where $E'$ is the electric field gradient. At our operating axial trap frequency of 210 kHz, this translates to variations in the quadrupole shift due to changes in the axial trap frequency of $27\pm12 \text{ mHz}/\text{kHz}$. The offset of $8.9(8)\text{ Hz}$ is due to the quadratic Zeeman shift, which agrees with the estimated value of 8 Hz for the applied magnetic field of 3.930 G. Any change in the magnitude of the applied magnetic field near our operating value of 3.930 G gives a variation of the quadratic Zeeman shift of $4 \text{ mHz}/\text{mG}$. Using the ions as a probe, we measure both the magnetic field and the axial trap frequency during the course of the experiment and correct for their contributions from the oscillation signal. Over the course of our 23-hour-long run, our axial trap frequency varies within $\sim 1~\text{ kHz}$ and the magnetic field within $1~\text{mG}$. These instabilities translate into variations of the correction for the quadrupole shift of $\sim30$~mHz and for the magnetic field of 3~mHz to the oscillation frequency. Fitting the model in Eq.~(\ref{eq:model}) to the corrections only, we find that not taking into account the axial frequency instability would cause a false Lorentz-violation signal with amplitudes of less than 3 mHz, while not correcting for the magnetic field instabilities would cause  a signal with amplitudes of less than 0.5 mHz. Thus, in principle no correction for their drift would have been necessary. We note also that by measuring those quantities during the measurement run, their contributions are expected to average down as fast as the primary measurement signal and thus should pose no limitation for improved Lorentz symmetry tests with longer measurement runs.

The oscillating electric field from the rf electrodes of the trap induces ac Stark shifts of the atomic transitions of the ions. The amplitude of the oscillating field experienced by the ions depends on the stray background static electric field. For our trap, we estimate that the stray electric field at the vicinity of the ions is $\sim$5 V/cm. This produces a differential ac Stark shift between the $\ket{\pm1/2}$ and $\ket{\pm5/2}$ states to be $\sim$120 mHz (Ref. \citenum{Yu1994}). The stability of the stray field is expected to be better than $10^{-2}$ level during the course of the experiment, which translates to less than 4 mHz change in the oscillation frequency for the two-ion state.

\subsection*{Statistical analysis of the data.}

After each measurement block as shown in Figure \ref{fig:measurement}b, we obtain a data point for the frequency difference between both states. We then bin the data points within 60 minutes intervals. The errorbar for each binned data point is assigned using the calculated standard deviation within each bin. To extract the amplitudes of Lorentz-violation, we perform a weighted least-square-fit of the binned data points to the model given in Eq.\ (\ref{eq:model}). We scale the 1$\sigma$ standard errors of the fitted parameters with $\sqrt{\chi^2_\text{reduced}} = 1.3$ to conservatively account for other remaining systematics.

\clearpage

\section*{Supplementary Information}

\subsection*{Calculation of the energy shift due to the Lorentz-violation for \Ca.}
Violations of Lorentz symmetry and Einstein's equivalence principle in bound electronic states
result in a small shift of the Hamiltonian that can be described by~\cite{HohLeeBud13}
\begin{equation}
\delta \mathcal{H}=-\left(  C_{0}^{(0)}-\frac{2U}{3c^{2}}c_{00}\right)
\frac{\mathbf{p}^{2}}{2}-\frac{1}{6}C_{0}^{(2)}T^{(2)}_{0},\label{eq1}
\end{equation}
where we use atomic units, $\mathbf{p}$ is the momentum of a bound electron, $U$ is the Newtonian potential, and $c$ is the speed of light. The parameters $C_{0}^{(0)}, C_{0}^{(2)}$ and $c_{00}$ are elements in the $c_{\mu\nu}$ tensor which characterises Lorentz-violation. The relativistic form of the $\mathbf{p}^2$ operator is $c\gamma_0\gamma^j p_j$ (a summation is implied by repeat indices), where $\gamma^i$ are the Dirac gamma matrices. The non-relativistic form of the $T^{(2)}_{0}$ operator is
$T^{(2)}_{0}=\mathbf{p}^{2}-3p_{z}^{2},$ where $p_z$ is the component of the momentum along the quantisation axis,
and the relativistic form is $T^{(2)}_{0}  = c\gamma_0\left(\gamma^j p_j-3\gamma^3 p_3\right)$.
Therefore, the shift of Ca$^+$ $3d~^2\text{D}_{5/2}$ energy level due to the $c_{\mu \nu}$ tensor depends on the values of
$\langle 3d~^2\text{D}_{5/2}|\mathbf{p}^{2}| 3d~^2\text{D}_{5/2}\rangle$ and $\langle 3d~^2\text{D}_{5/2}|T^{(2)}_0|3d~^2\text{D}_{5/2} \rangle$ matrix elements.

Using the Wigner-Eckart theorem we express the matrix element of the irreducible tensor operator $T_0^{(2)}$ through the reduced matrix element of the operator $T^{(2)}$ as
\begin{equation}
\label{eq10}
\langle Jm_J|T^{(2)}_0|Jm_J \rangle = \frac{-J\left( J+1\right)  +3m_J^{2}}{\sqrt{\left(
2J+3\right)  \left(  J+1\right)  \left(  2J+1\right)  J\left(  2J-1\right)}} \,
\langle J||T^{(2)}||J \rangle.
\end{equation}
The expressions for the $\mathbf{p}^{2}$ and
$T^{(2)}$ matrix elements are given in the supplementary material of ref.\ \citenum{HohLeeBud13}.
The values of angular factors in Eq.\ (\ref{eq10}) are $-0.27951+0.22361~ m_J^2$ for $3d~^2\text{D}_{3/2}$
and $-0.21348+0.073193~ m_J^2$ for $3d~^2\text{D}_{5/2}$.

First, we calculated the required matrix elements in a lowest-order Dirac-Fock (DF) and then including random-phase approximation (RPA).
Next, we carry out  much more accurate calculations using the configuration interaction method with single and double excitations (CI-SD) and
four variants of the all-order (linearised coupled-cluster) method\cite{SafJoh08}.
The virial theorem is also used for the $\mathbf{p}^2$ calculations.

The results are summarised in Table~\ref{tabs}. We note that we list the \textit{reduced} matrix elements for the $T^{(2)}$ operator
but actual matrix elements for the $\mathbf{p}^{2}$ operator because there is no necessity to introduce
reduced matrix elements for a scalar operator.
The values in the DF(FC) and DF columns are lowest-order DF values calculated with and without the \textit{frozen core} approximation.
In the \textit{frozen  core} approximation the DF equations for the core electrons are solved self-consistently first and the valence
orbital is calculated with unchanged, \textit{i.e.} ``\textit{frozen}'' core. For the $\mathbf{p}^2$ operator such approximation appears to give
very poor results for the $3d$ states. If the core orbitals are allowed to vary together with the valence orbital, the lowest-order value
is only 16\% away from the final virial theorem value. Addition of the RPA correction to the frozen-core DF value fixes this problem as well, as RPA corrections describe reaction of the core electrons to an externally applied perturbation. The perturbation produced by the operator ${\bf p}^2$ is very large and, as a result,
the RPA corrections for $ \langle \psi |\mathbf{p}^{2}| \psi \rangle$ matrix elements are large.
Such problem does not arise for the $T^{(2)}$ operator; the correlation correction to its matrix elements is much smaller and the
accuracy of the resulting values is much higher.

The  CI-SD calculations are carried out using the Dirac-Fock basis for the occupied core and
valence atomic states and DF-Sturm basis for unoccupied
virtual orbitals; the frozen-core approximation is not used. The description of the DF-Sturm equations is given in
ref.\ \citenum{Tupitsyn_05,Tupitsyn_10}.
The configuration state functions (CSF) are
constructed from the one-electron wave functions
as a linear combination of Slater determinants. The set of the CSFs
is generated including all single and double excitations into
one-electron states of the positive spectrum. Single excitations are allowed to all core shells,
double excitations are allowed to $3s$ and $3p$ core shells.
\begin{table*}
\caption{\label{tabs} Lowest-order DF, DF+RPA, CI+single-double excitations (CI-SD), and all-order  results for the
$\langle 3d~^2\text{D}_{J}|p^2|3d~^2\text{D}_{J}\rangle$ and $\langle3d~^2\text{D}_{J}||T^{(2)}_0||3d~^2\text{D}_{J}\rangle$ matrix elements in Ca$^+$ in atomic units. The virial
theorem values are listed in the column ``VT''. The values in the DF(FC) and DF
columns are lowest-order DF values calculated with and without the \textit{frozen core} approximation.}
\begin{ruledtabular}
\begin{tabular}{lcccccccc}
Matrix element                               & DF(FC)  & DF   & RPA  &  CI+SD  &All-order &  VT   & Final  \\ \hline
$\langle 3d~^2\text{D}_{3/2}|p^2|3d~^2\text{D}_{3/2}\rangle$       & 3.05    & 0.67 & 0.66 &  0.73   &0.83      & 0.748 & 0.75(9)\\
$\langle 3d~^2\text{D}_{5/2}|p^2|3d~^2\text{D}_{5/2}\rangle$       & 3.04    & 0.66 & 0.66 &  0.73   &0.83      & 0.748 & 0.75(9)\\
$\langle 3d~^2\text{D}_{3/2}||T^{(2)}||3d~^2\text{D}_{3/2}\rangle$ & 5.45    & 6.22 & 5.72 &  6.89   &7.09      &       & 7.09(12)\\
$\langle 3d~^2\text{D}_{5/2}||T^{(2)}||3d~^2\text{D}_{5/2}\rangle$ & 7.12    & 8.11 & 7.47 &  8.98   &9.25      &       & 9.25(15)\\
\end{tabular}
\end{ruledtabular}
\end{table*}
To calculate the value $\langle v|\mathbf{p}^2|v \rangle$, we also used the approach
based on the virial theorem. In the nonrelativistic limit the virial
theorem can be written in the form
$$
E = - \frac{1}{2} \, \langle \Psi | \sum_i \mathbf{p}^2(i) | \Psi \rangle \,,
$$
where $E$ is a total energy of the system. Therefore,  the value
$\langle v|p^2|v \rangle$ can be calculated using the removal energies of the valence electron.
The virial theorem gives us a possibility to calculate the expectation value of the $\mathbf{p}^2$ operator
as a difference of the total energies $E_N$ and $E_{N-1}$ of $N$ and $N-1$ systems multiplied by 2. Since the differential energy $E$
can be calculated with an accuracy much higher than the wave function $\Psi$,
this approach is appropriate for the light atoms and ions where relativistic
effects are negligible. The virial theorem results that use experimental data for the $3d$ removal energies from ref.\ \citenum{NIST} are listed in the column ``VT''.

We have also carried out the calculations of the $\langle\Psi|\mathbf{p}^{2}|\Psi\rangle$ and $\langle\Psi||T^{(2)}||\Psi\rangle$
matrix elements using the all-order (linearised coupled-cluster) method\cite{SafJoh08}. The all-order method gave very accurate
values of the $3d_j$ lifetimes~\cite{KreBecLan05} and quadrupole moments~\cite{JiaAroSaf08} in a Ca$^+$ ion.
In the  all-order method, single, double, and partial triple excitations of Dirac-Hartree-Fock wave functions are included to all orders
of perturbation theory. We refer the reader to the review\cite{SafJoh08} for the description of the all-order method and its applications.
Both single-double (SD) and single-double-partial triple (SDpT) \textit{ab initio} all-order calculations were carried out. In addition,
a  scaling of the dominant terms\cite{SafJoh08} was carried out for both SD and SDpT calculations to improve the accuracy and to evaluate
the uncertainty of the final values. The calculations were carried out with both nonrelativistic and relativistic  operators;
the differences were found to be negligible at the present level of accuracy. The values calculated with relativistic operators are listed
in Table~\ref{tabs}.

The virial theorem values are taken as final for the matrix element of the $\mathbf{p}^2$ operator. The uncertainty of 12\% is estimated
as the difference of the virial theorem and all-order values.
The SD scaled values are taken as final for the $T^{(2)}$ operator (see ref. \citenum{KreBecLan05,JiaAroSaf08} for the discussion of the choice
of the final all-order values). The uncertainty is determined as the spread of the four all-order values.
Substituting the final all-order values of the $\langle3d~^2\text{D}_{J}||c\gamma_0\left(\gamma^j p_j-3\gamma^3 p_3\right)||3d~^2\text{D}_{J}\rangle$
matrix element into Eq.~(\ref{eq10}) and using virial theorem value of $\langle3d~^2\text{D}_{J}|\mathbf{p^2}|3d~^2\text{D}_{J}\rangle$ we get:
 \begin{align}
\mathrm{3d}~^2\text{D}_{3/2}  :\frac{\triangle E}{h}&\approx-2.46\times
10^{15}\left(  C_{0}^{(0)}-\frac{2U}{3c^{2}}c_{00}\right)
 +\left(  2.17\times10^{15}-1.47\times10^{15}\text{\thinspace}m_J^{2}\right)
C_{0}^{(2)},\\
\mathrm{3d}~^2\text{D}_{5/2}  :\frac{\triangle E}{h}&\approx-2.46\times
10^{15}\left(C_{0}^{(0)}-\frac{2U}{3c^{2}}c_{00}\right)
 +\left(  2.16\times10^{15}-7.42\times10^{14}\text{\thinspace}m_J^{2}\right)
C_{0}^{(2)},
\end{align}
where the uncertainty of the coefficients standing in front of the $\left(C_{0}^{(0)}-\frac{2U}{3c^{2}}c_{00}\right)$ and  $C_{0}^{(2)}$
terms are estimated to be 12\% and 2\%, respectively.
The atomic units are converted to Hz using $1$ a.u. $\approx h\cdot\left(  6.57968\times10^{15}\text{ Hz}\right)$,
where $h$ is Planck constant.

The frequency difference (in Hz) between the shifts of the $m_J=5/2$ and $m_J=1/2$ states for a pair of \Ca~used in our experiment is given by
\begin{align}
2\times \frac{1}{h}\left( E_{m_J=5/2} -E_{m_J=1/2}\right) &= -1.484\times10^{15} \textrm{ Hz } \left((5/2)^2-(1/2)^2\right)
\cdot C_{0}^{(2)}\\
&=-8.9(2)\times10^{15}~ \textrm{ Hz } \cdot C_{0}^{(2)}.
\end{align}

\clearpage

\subsection*{Transformation of the $c_{\mu\nu}$ tensor from the laboratory frame to the Sun's rest frame.}

As shown in the previous section, the energy shift of a single \Ca~between the $\ket{m_J=5/2}$ and $\ket{m_J=1/2}$ states in the $^2\text{D}_{5/2}$ manifold is given by
\begin{eqnarray}
\Delta E_\text{LLI} = \mathcal{Q}\cdot C^{(2)}_0,
\end{eqnarray}
where $\mathcal{Q} = -4.45(9)\times 10^{15}\text{ Hz}$, $C_0^{(2)} \equiv c_{jj}-3c_{33}$, with summations implied by repeated indices, contains elements in the $c_{\mu\nu}$ tensor in the local laboratory frame.

Because of the Earth's motion, $c_{\mu\nu}$ in the local laboratory frame varies according to the time-dependent Lorentz transformation given by
\begin{align}
c_{\mu\nu} = c_{MN}\Lambda^M_\mu\Lambda^N_\nu,
\end{align}
where $\Lambda$ is the Lorentz transformation matrix and $c_{MN}$ is $c_{\mu\nu}$ written in the Sun-centred, celestial-equatorial frame (SCCEF). The matrix $\Lambda$ consists of a rotation and a velocity boost of the experiment with respect to the Sun. In the laboratory frame, we define the $\hat{x}$ axis to point to the East, $\hat{y}$ axis to point to the North and $\hat{z}$ axis to point upward. The rotation matrix that transforms from the SCCEF to the local laboratory frame is given by
\begin{align}
R =\left( \begin{array}{ccc}
 -\sin(\omega_\oplus T) & \cos(\omega_\oplus T) &0\\
 -\cos\chi\cos(\omega_\oplus T) & -\cos\chi\sin(\omega_\oplus T) & \sin\chi\\
 \sin\chi\cos(\omega_\oplus T) & \sin\chi\cos(\omega_\oplus T) & \cos\chi\end{array} \right),\label{SI:eq:rotation_matrix}
\end{align}
where the angle $\chi\sim52.1$\textdegree~is the colatitude of the experiment (Berkeley, CA), $T$ is time since vernal equinox of 2014 and $\omega_\oplus = 2\pi/23.93$~h\ is the sidereal angular frequency of the Earth's rotation. The boost of the experiment in the SCCEF is given by
\begin{align}
\vec{\beta} =\left( \begin{array}{c}
-\beta_\oplus \sin(\eta)\cos(\Omega T)\\
\beta_\oplus \cos(\eta)\cos(\Omega T)-\beta_L\sin(\chi)\cos(\omega_\oplus T) \\
-\beta_\oplus \sin(\Omega T)+\beta_L\sin(\chi)\sin(\omega_\oplus T) 
 \end{array} \right),\label{SI:eq:boost_matrix}
\end{align}
where $\beta_\oplus\sim10^{-4}$ is the boost from the Earth's orbital velocity and $\beta_L\sim1.5\times10^{-6}$ is the boost from the Earth's rotation, $\Omega$ is the yearly sidereal angular frequency and $\eta\sim23.4$\textdegree~is the angle between the ecliptic plane and the Earth's equatorial plane.

The parameter relevant to our experiment is $C_0^{(2)}$. With the Lorentz transformation applied to $c_{\mu\nu}$ in the SCCEF, we can write the value of $C_0^{(2)}$ in the local laboratory frame in terms of $c_{\mu\nu}$ in the SCCEF to be
\begin{align}
C_0^{(2)} = A +\sum_j\left(C_j\cos(\omega_j T) + S_j\sin(\omega_j T)\right),
\end{align}
where $C_j$, $S_j$ and $\omega_J$ are amplitudes and angular frequency given in Table \ref{SI:table:coeff}, and $A$ is a constant offset. For our 23-hour measurement, the leading order of the time-dependent Lorentz-violation signal is given by
\begin{align}
C_0^{(2)} &= -3\sin(2\chi)c_{XZ}\cos(\omega_\oplus T) - 3\sin(2\chi)c_{YZ}\sin(\omega_\oplus T)-\nonumber\\
& \qquad -\frac{3}{2} (c_\text{XX}-c_\text{YY}) \sin ^2(\chi )\cos(2\omega_\oplus T)-3 c_{XY} \sin ^2(\chi )\sin(2\omega_\oplus T).
\end{align}
We fit our binned 23-hour measurement data to this model and extract Lorentz-violation parameters, where we report in Table \ref{table:cparam} uncorrelated combinations of parameters by diagonalising the covariance matrix from the fit. We scale the $1\sigma$ uncertainties from the fit with $\sqrt{\chi^2_\text{reduced}} = 1.3$ to conservatively account for other remaining systematics.

With a year-long measurement, we expect to reach the sensitivity in the ions oscillation frequency of 1 mHz. This level of sensitivity allows us to bound $c_{TX}$, $c_{TY}$ and $c_{TZ}$ at the $10^{-16}$ level, which will improve the current limits\cite{HohLeeBud13} for these parameters for at least an order of magnitude.  

\begin{table*}
\caption{Amplitudes of various frequency components for $C_0^{(2)}$ expressed in terms of $c_{\mu\nu}$ in the SCCEF. The frequencies $\omega_\oplus$ and $\Omega$ are the daily and yearly sidereal angular frequency, respectively. The angle $\chi\sim52.1$\textdegree~is the colatitude of the experiment (Berkeley, CA). The angle $\eta\sim23.4$\textdegree~is the angle between the ecliptic plane and the Earth's equatorial plane. $\beta_\oplus\sim10^{-4}$ is the boost from the Earth's orbital velocity and $\beta_L\sim1.5\times10^{-6}$ is the boost from the Earth's rotation. For our 23-hour-measurement, contributions from these two boosts are negligible.}
\label{SI:table:coeff}
\begin{ruledtabular}
\begin{tabular}{lcccc}
$\omega_j$                   &\vline& $C_j$  &\vline& $S_j$ \\ \hline
$\omega_\oplus$ &\vline& $-3\sin(2\chi)c_{XZ}+2c_{TY}\beta_L$&\vline&$-3\sin(2\chi)c_{YZ}-2c_{TX}\beta_L$\\
$2\omega_\oplus$ &\vline&$-\frac{3}{2} (c_\text{XX}-c_\text{YY}) \sin ^2(\chi )$ &\vline&$-3 c_{XY} \sin ^2(\chi )$ \\
$\Omega$ &\vline&$-\frac{1}{2} \beta_\oplus (3 \cos (2 \chi )+1) (c_\text{TY} \cos (\eta )-2
   c_\text{TZ} \sin (\eta ))$ &\vline& $\frac{1}{2} \beta_\oplus c_\text{TX} (3 \cos (2 \chi )+1)$\\
$2\Omega$ &\vline&0 &\vline& 0\\
$\Omega-\omega_\oplus$ &\vline&$\frac{3}{2} \beta_\oplus c_\text{TX} \sin (\eta ) \sin (2\chi )$ &\vline& $-\frac{3}{2} \beta_\oplus \sin (2\chi )
   \left(c_\text{TY} \sin \left(\eta\right)+c_\text{TZ}(1+\cos
   \left(\eta \right))\right)$\\
$\Omega+\omega_\oplus$ &\vline& $\frac{3}{2} \beta_\oplus c_\text{TX} \sin (\eta ) \sin (2\chi )$&\vline& $-\frac{3}{2} \beta_\oplus  \sin (2\chi )
   \left(c_\text{TZ} (1-\cos(\eta))-c_\text{TY} \sin
   (\eta)\right)$\\
$2\Omega-\omega_\oplus$ &\vline&0 &\vline& 0\\
$2\Omega+\omega_\oplus$ &\vline& 0&\vline& 0\\
$\Omega-2\omega_\oplus$ &\vline&$-3 \beta_\oplus c_\text{TY} \cos ^2\left(\frac{\eta }{2}\right) \sin ^2(\chi
   )$ &\vline& $-3 \beta_\oplus c_\text{TX} \cos ^2\left(\frac{\eta }{2}\right) \sin ^2(\chi
   )$\\
$\Omega+2\omega_\oplus$ &\vline& $3 \beta_\oplus c_\text{TY} \sin ^2\left(\frac{\eta }{2}\right) \sin ^2(\chi )$&\vline& $-3 \beta_\oplus c_\text{TX} \sin ^2\left(\frac{\eta }{2}\right) \sin ^2(\chi
   )$\\
$2\Omega-2\omega_\oplus$ &\vline&0 &\vline& 0\\
$2\Omega+2\omega_\oplus$ &\vline& 0&\vline& 0\\
\end{tabular}
\end{ruledtabular}
\end{table*}

\end{document}